\documentclass[12pt,aps,prb]{revtex4}   % style for Physical Review B and AJP are similar

\usepackage{amsmath}    % need for subequations
\usepackage{graphicx}   % for figures

\begin{document}

\title{Echoes from the Moon}
\author{Luca Girlanda}
 \affiliation{INFN, Sez. di Pisa, Largo Filippo Buonarroti, I-56127
 Pisa, Italy\\and Liceo Scientifico ``E. Fermi'', Via
 Enrico Fermi 2, I-54100 Massa, Italy }
 \email{girlanda@pi.infn.it}   %optional
\date{\today}

\begin{abstract}
We report on a determination of the Earth-Moon distance performed by
students of an Italian high school, based on measurements of the
time delay of the ``echo'' in the radio communications between
Nasa mission control in Houston and the Apollo astronauts on the lunar surface.
By using an open-source audio-editing software, the distance can be
determined with three digits accuracy, allowing to detect  the
effect due to the eccentricity of the orbit of the Moon.
\end{abstract}

\maketitle
\section{Introduction}
How far is the Moon? A possible answer to this question can be given
knowing the Moon diameter, which may be found to be about 2/7 of the
Earth diameter from the duration of a lunar eclipse,
cfr. Refs.~\cite{rogers,cowley,bruning}. Dividing by the angular
width, about half a degree, one finds about 60 times the Earth radius,
a result which is considered to lie at the 
foundation of the Newtonian inverse-square law of gravitation, since the
lunar centripetal acceleration is found to be 3600 times smaller than
$g$.  

Nowadays the Earth-Moon distance $d_{\mathrm{EM}}$ is constantly
monitored, since the first experiment on August 1st 1969
\cite{lunarranging}, by 
measuring the 
time of flight (about 2.5~s) of a laser pulse which is directed from Earth
towards an optical retroreflector array placed on the lunar surface
during the Apollo 11 mission. In this note we report on a
measurement based on the same principle, 
using instead the delay in the communications between  NASA mission control in Houston
and the Apollo astronauts on the lunar surface. This activity was
carried out by students of an Italian high-school (age range
14-19), by analyzing  
conversations between the astronauts and mission control. These
conversations were recorded by NASA and are
 available from NASA's web site \cite{apollo}.

Students were divided in 10 groups of two or three. A first round of
measurements was performed with chronometers, by listening to the conversation with
Neil Armstrong during the Apollo 11 mission, during which the famous sentence ``one
small step for man, one giant leap for mankind'' can be heard.
The 10 groups  were provided with the tapescripts and they measured the delays in
Houston's and Armstrong's replies, as shown in
Table~\ref{tab:chrono}. Only the delays in Armstrong's replies 
  are affected by the time of flight of the radio signal, since the
  tape was recorded at Houston. From the minimum delay in Armstrong's
  replies (last column of the 2nd row) an upper bound for 
the Earth-Moon distance was 
found, $d_{\mathrm{EM}} < (4.5 \pm 0.7) \cdot 10^8$~m. 
\begin{table}[h]
\begin{tabular}{|l|c|c|c|c|c|c|c|}
\hline
Replies from  & \multicolumn{7}{|c|}{Time delays (s)}\\
\hline
Houston   & $1.55\pm0.15$ & $0.35 \pm 0.15$ & &
 $1.35 \pm 0.25$ & $1.7\pm 0.2$ & $0.85 \pm 0.15$ & \\
\hline
Armstrong &               &                 &  $4.05\pm
 0.25$ &         &              &                 & $3.0 \pm 0.2$ \\   
\hline
\end{tabular}
\caption{Time delays of the replies in the 3-minutes conversation between Houston and
  Armstrong during which the famous sentence ``one small step for man, one giant
  leap for Mankind'' can be heard. The errors represent the ranges of
  values measured by the 10 groups of students with
  chronometers. The very short delay in the 2nd column corresponds to a
  radio check requested by Armstrong and promptly replied by Houston.
 The mp3 file of this famous conversation is
  available at the  
  NASA web site \cite{nasa_conv}. \label{tab:chrono}}
\end{table}

It came as a surprise that sometimes a clear
echo of the sentences from the Earth was audible, due to the retransmission of
the signal from the speaker  through Armstrong's microphone. In such cases a much more
accurate measurement of the time of flight of the radio signals is possible with the help of 
audio-editing software.\footnote{The same
  observation was made in Ref.~\cite{keeports} and 
  used to estimate the speed of light, see also Ref.~\cite{glick}.}  
 This allows for a very
precise determination of $d_{\mathrm{EM}}$.

\section{Data analysis}
\begin{figure}
\centerline{\includegraphics[width=12cm]{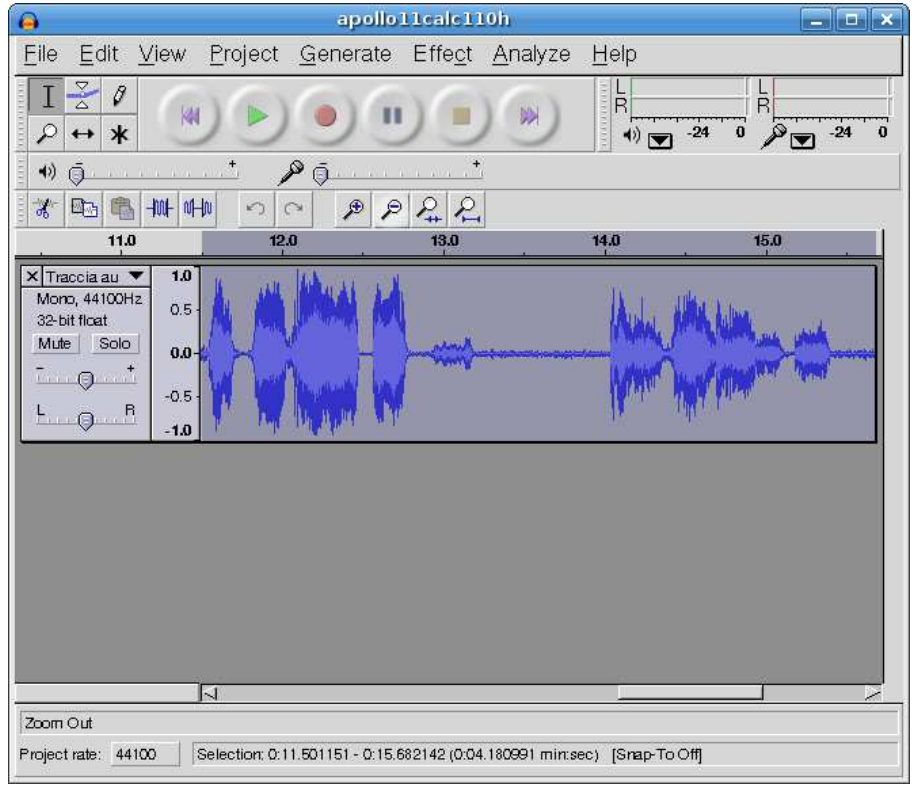}}
\caption{Audio output of the words ``... Houston, AOS, over'', as
  displayed using the opensource software Audacity \cite{audacity}; the
  pattern clearly reveals the echo effect. The small signal at about
  13~s is a beep (``Quindar tone'') used to trigger ground station
  transmitters. The time scale can be arbitrarily expanded allowing an
  accurate determination of the echo delay. \label{fig:over}}  
\end{figure}

We have used Audacity \cite{audacity}, a freely available
open-source program, running under Windows, Mac OS X and GNU/Linux
operating systems. 
This software allows to visualize the level of audio output as function of
time, with the time scale arbitrarily adjustable. In this way 
the time windows of single syllables and of their echoes can be clearly
identified and isolated. We have singled out the word ``over'' in the
sentence ``Columbia, Columbia, this is Houston, AOS [acquisition of
  signal], over'', at 110h:07m:58s from ignition time, for which a clear echo was
audible, cfr. Fig.~\ref{fig:over}.

 Several methods of measurement of the time delay of the echo can be
adopted and the students were given freedom to devise the more appropriate 
one: some chose to mark the times at the beginning or at the end of the
syllables, others at the maximum of the signal within a syllable. Each
measurement has an  associated uncertainty, that the students were asked to
estimate as well. For instance, the duration of a syllable and of its echo can
be measured,  
and the difference be taken as an estimate of the uncertainty. 
Specifically, four groups chose to use the beginning of the word ``over'',
three the end of the same word, and three the maximum of the audio signal
within the word, the uncertainties associated to the latter method
being however substantially larger.
As a final result, without affording a thorough statistical analysis,
students agreed to 
take the average of the 10 measurements with the maximum shift from average as 
uncertainty, $\Delta t = (2.620 \pm 0.007)$~s. 
This translates into $d_{\mathrm{EM}}=(3.93\pm 0.01)\cdot  10^8$~m,  quite an
accurate measurement. 
The above error estimate is clearly larger than what would be obtained by a
correct statistical treatment of the data: the standard deviation of
the 10 measurements is $\sim 0.005$~s, and therefore the predicted standard
deviation of the mean is $\sim 0.002$~s. A maximum likelihood fit to
the 10 measurements would yield instead $\Delta t = (2.616\pm
0.001)$~s,
where the central value is the weighted average and the error
corresponds to an increase of the $\chi^2$ by 1;  correspondingly,
 $d_{\mathrm{EM}}=(3.921\pm 0.002)\cdot  10^8$~m. 

Which number one has to compare to? At this
level of accuracy one should be able to detect several effects, such
as the variation of $d_{\mathrm{EM}}$  due
to the ellipticity of the orbit (up to $5\cdot 10^6$~m per day). This has been the second task of our activity.
To this aim we have chosen to analyze the Apollo~17 mission, which
lasted longer than all the others, about 300 hours.
Special attention had to be payed to the fact that even the Earth
rotation would affect the measurements. One way to isolate the 
effect due to the eccentricity of the orbit is to take measurements at
24h-separated times (or rather at 24h50'-separated times): at
intermediate times radio signals were probably following different paths.

\section{Results}
\begin{figure}
\centerline{\includegraphics[width=12cm]{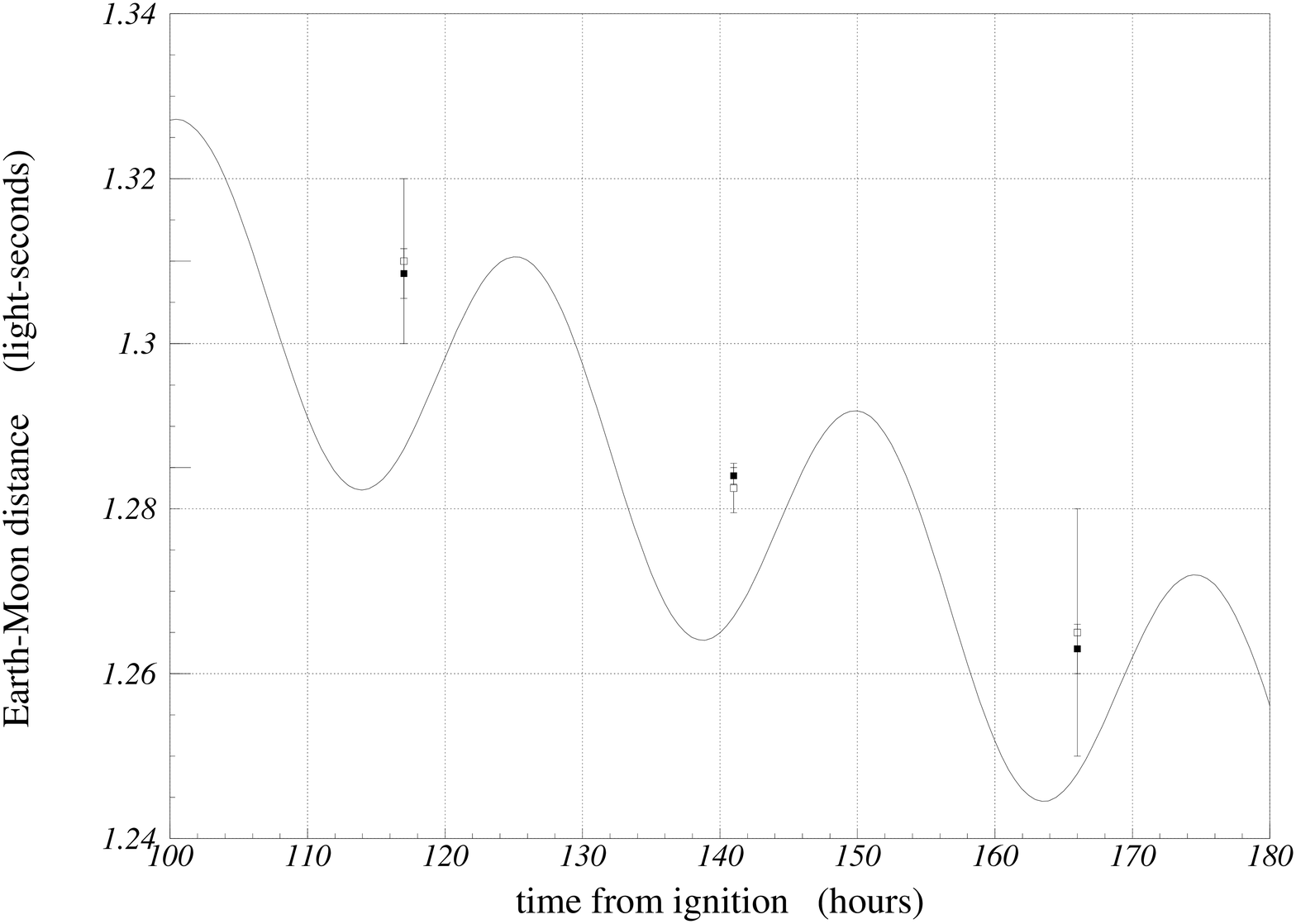}}
\caption{Distance between Houston and the astronauts on the Moon in
  light-seconds. Data points 
  correspond to our measurements, i.e. the values in
  Table~\ref{tab:measurements} divided by~2. Empty squares are the
  students' results (3rd column of the Table), filled squares (with
  much smaller error bars) are from
  the 4th columns. The curve is obtained by
  using the Moon ephemerides \cite{jpl} and refers to the distance
  between Houston and the Moon center.\label{fig:data}}
\end{figure}

We have identified three sentences with a clear
enough echo,  which at the same time almost satisfied the 24h separation
constraint.  Our measurements of the minimum time delay of the echoes are
reported in Table~\ref{tab:measurements}.  
\begin{table}[h]
\begin{tabular}{|l|c|c|c|c|}
\hline
word & time (hh:mm:ss) & \multicolumn{2}{|c|}{delay (s)}& time shift (s)\\
\hline
``Houston'' & 117:03:11 & $2.62 \pm 0.02$ & $2.617 \pm 0.006$ & $0.043
\pm 0.006$\\
``Geno''    & 141:24:11 & $2.565 \pm 0.006$& $2.568 \pm 0.002$ &$
0.034 \pm 0.002$ \\
``three''   & 166:06:22 & $2.53 \pm 0.03$&  $2.526 \pm 0.006$ & $
0.030 \pm 0.006$\\
\hline
\end{tabular}
\caption{Time delays of the echoes during the
  radio communications of the Houston mission control with the Apollo 17 astronauts on
  the Moon. The delays were measured by 10 groups of students. In the
  3rd column the average of the 10 measurements is shown, with the error
  representing the 
  maximum deviation from the average. A more proper statistical treatment
  of the data (maximum likelihood fit, as explained in the text) would yield the results in the
  4th column.   The times in the 2nd column are counted from ``ignition
  time''. In the 5th column we show  the differences of the times in the
  4th column with the Moon ephemerides \cite{jpl}. The mp3 files of 
  the registration are available at the 
  NASA web site \cite{nasa_conv}. \label{tab:measurements}}
\end{table} 
The corresponding values of $d_{\mathrm{EM}}$ are plotted in
  Fig.~\ref{fig:data}.
 
Also shown for comparison are the lines from the Moon
  ephemerides, calculated with the JPL HORIZONS system \cite{jpl},
  setting as location the 
  geographical coordinates of Houston, latitude $29^\circ   45'26''$~N,
  longitude $95^\circ 21'37''$~W; to this aim one can  also 
  profitably use the opensource program Stellarium
  \cite{stellarium}, which yields the same results and is much more
  fun for students. The
  zero of the time axis is the ``ignition time'' for the 
  Apollo 17 mission, 5:33:00 a.m. (GWT) on December 7th 1972.

  It can clearly be seen that our measurements follow the decrease of the
  Earth-Moon distance, with a shift of about  $0.017 \pm 0.001$~s (half of the
  values in the 5th column of Table~\ref{tab:measurements}), to be
  ascribed to the delay in electronics and other systematic
  effects. One such effect is the fact that the radio signal did not
  follow a straight line from Houston to the Moon surface. Indeed, according
  to the report on the Apollo radio communication system \cite{usb},
  three main ground antenna were used during 
  the lunar phases of the missions, which were situated in Goldstone
  (California), Canberra (Australia) and Madrid (Spain), the closest one
  to Houston being 2100~Km far. This produces a delay of
  at least 0.007~s. An additional delay, possibly of the same order of
  magnitude, could be due to the fact that the radio signal was probably
  transmitted from the receiving antenna to Houston through a
  satellite. It should also be noted that the Moon ephemerides give 
  the distance of the center of the Moon from Houston, while the
  astronauts were on the lunar surface (at about $\sim 20^\circ$ North
  latitude, $\sim 30^\circ $ East longitude \cite{report}), which is
  $\sim 1500$~Km closer to Houston than the center of the Moon. 
  All these effects, which in some case, however, tend to compensate
  one-another, are of the same order of magnitude as the observed
  shift.

\section{Discussion}
The experiment that we have reported represents an exemple of ``open''
research-oriented activity, in which no ``correct answer''  can be
anticipated {\em a priori}.  This aspect should be emphasized whenever
possible in the teaching of physics, since it gives students the flavor of what
physicists do in their experimental or theoretical work. 
Another aspect of our experiment - common to most of
present-day experimental research in physics - is the analysis of raw
data (in our case the audio registrations) by means of sophisticated
software, which gives the opportunity to extend the
discussion on the ``error sources''.

The students have very much appreciated  the use of open-source software that
they could easily install in their computers at home, especially the
program ``Stellarium'', which simulates the appearance of the sky at
all times and from every location. 

In summary, the reported activity has constituted an important
cultural achievement for the students: for interdisciplinary reasons
-the study of the Apollo missions, one of the most fascinating
accomplishements of mankind, 
the use of English as a foreign language - but also for reasons more
proper to physics - the ellipticity of Kepler's orbits, the invariance
of the speed-of-light as foundation of the experiment, the propagation
of sound and of electromagnetic waves. Not to mention the thrill for
measuring a shorter time delay than allowed, which would constitute a
planetary scoop as evidence in favor of  much popular ``conspiracy
theories'' about  Apollo missions.

\begin{acknowledgments}
It is a pleasure to thank Prof.~L.~Bracci for a careful reading of the
manuscript, Prof.~M.~Coluccini and
Dr.~U.~Penco for 
inspiring discussions, the students of 1SD and 1SE (year 2005/2006) of Liceo
Scientifico ``A. Vallisneri'' (Lucca, Italy) for their involvement and
enthousiasm, and an anonymous Referee for helpful suggestions. This
activity was performed within the ``Progetto Lauree 
Scientifiche'', of the Italian Ministery of Education. 
\end{acknowledgments}

\end{document}